\documentclass{emulateapj}

\usepackage{natbib}
\usepackage{graphicx}

\bibpunct{(}{)}{;}{a}{}{,}   % to follow the ApJ style

\newcommand{\grb}{GRB\,030329}
\newcommand{\etal}{et~al.}

\newcommand{\gcn}{GCN Circular}

\newcommand{\Fm}{\ensuremath{F_{\mathrm{m}}}}
\newcommand{\vc}{\ensuremath{\nu_{\mathrm{c}}}}
\newcommand{\vm}{\ensuremath{\nu_{\mathrm{m}}}}
\newcommand{\va}{\ensuremath{\nu_{\mathrm{a}}}}
\newcommand{\tj}{\ensuremath{t_{\mathrm{j}}}}
\newcommand{\tnr}{\ensuremath{t_{\mathrm{nr}}}}
\newcommand{\Eiso}{\ensuremath{E_{\mathrm{iso}}}}
\newcommand{\Ecor}{\ensuremath{E_{\mathrm{cor}}}}
\newcommand{\ee}{\ensuremath{\epsilon_{\mathrm{e}}}}
\newcommand{\eb}{\ensuremath{\epsilon_{\mathrm{B}}}}
\newcommand{\vcg}{\ensuremath{\nu_{\mathrm{c,13}}}}

\slugcomment{Accepted for publication in ApJ}

\shorttitle{The radio afterglow of \grb}
\shortauthors{Van der Horst~\etal}

\begin{document}

\title{The radio afterglow of \grb\ at centimeter wavelengths: evidence for a structured jet or non-relativistic expansion}

\author{A.~J. van der Horst\altaffilmark{1}, E.~Rol\altaffilmark{2,1}, R.~A.~M.~J.~Wijers\altaffilmark{1}, R.~Strom\altaffilmark{3,1}, L.~Kaper\altaffilmark{1}, C.~Kouveliotou\altaffilmark{4}}

\altaffiltext{1}{Astronomical Institute, University of Amsterdam, Kruislaan 403, 1098 SJ Amsterdam, The Netherlands}
\altaffiltext{2}{Department of Physics and Astronomy, University of Leicester, University Road, Leicester LE2 7RH, United Kingdom}
\altaffiltext{3}{ASTRON, P.O. BOX, 7990 AA Dwingeloo, The Netherlands}
\altaffiltext{4}{NASA/MSFC, USRA at NSSTC, SD-50, 320 Sparkman Drive, Huntsville, AL 35805}

\begin{abstract}

We present our centimeter wavelength (1.4, 2.3 and 4.8 GHz) light curves of the afterglow of \objectname[GRB 030329]{\grb}, which were obtained with the Westerbork Synthesis Radio Telescope. 
Modeling the data according to a collimated afterglow results in a jet-break time of 10 days. This is in accordance with earlier results obtained at higher radio frequencies. However, with respect to the afterglow model, some additional flux at the lower frequencies is present when these light curves reach their maximum after 40-80 days. We show that this additional flux can be modeled with two or more components with progressively later jet breaks. From these results we infer that the jet is in fact a structured or a layered jet, where the ejecta with lower Lorentz factors produce additional flux that becomes visible at late times in the lowest frequency bands. We show that a transition to non-relativistic expansion of the fireball at late times can also account for the observed flux excess, except for the lowest frequency (1.4 GHz) data.

\end{abstract}

\keywords{gamma rays: bursts -- radio continuum: general -- radiation mechanisms: non-thermal}

%\maketitle

%
%________________________________________________________________

\section{Introduction \label{SECTION:GRB030329-INTRODUCTION}}

Since the discovery of afterglow emission of gamma-ray bursts (GRBs) at X-ray, optical and radio wavelengths \citep{costa1997:nature387,vanparadijs1997:nature386,frail1997:nature389}, it has become clear that broad-band observations are needed to determine the physical processes producing the afterglow emission in the context of the available models, the most popular being the fireball model \citep[e.g.,][]{rees1992:mnras258,meszaros1997:apj476}. Obtaining the overall shape of the energy distribution and the time evolution of the GRB afterglow provides information about the intrinsic energy, both in electrons and in magnetic fields, as well as about the matter into which the GRB blasted its ejecta \citep[see, e.g.,][]{wijers1999:apj523}. Although optical observations alone can constrain the value of some of these physical parameters, observations covering the radio to X-ray wavelength regions are required to determine all of them.

The self-absorption frequency, \va, of the afterglow broad-band spectrum can often be constrained by radio observations at centimeter wavelengths \citep[e.g.,][]{wijers1999:apj523}. As the afterglow spectrum evolves, the two other characteristic break frequencies in its broad-band spectrum (the frequency at the peak flux, \vm, and the cooling frequency, \vc) enter the radio regime as well, although in practice the flux level at \vc\ is below the detection limit of current radio telescopes. The latter frequency can usually be determined from the available optical and X-ray data, which span the frequency range where the cooling frequency is found during the first days of the afterglow. These break frequencies and their time evolution uniquely determine the parameters that make up the fireball model and its evolution.

GRB\,030329 is the closest gamma-ray burst discovered so far for which an afterglow has been found\footnote{GRB\,980425/SN\,1998bw at $z = 0.0085$ was closer, but no afterglow was found}. At a redshift of $z = 0.1685$ \citep{greiner2003:gcn2020}, its afterglow was discovered in $R = 12.4$ magnitude, just 67 minutes after the GRB itself \citep{sato2003:apj599}, about 100 times brighter than the average GRB afterglow. The brightness of the afterglow made it possible to study its evolution for a long time and in detail over a broad range of frequencies, from X-ray to centimeter wavelengths. Furthermore, its proximity provided an excellent opportunity to look for a supernova signature in both the light curve and the spectrum, as predicted by the collapsar model \citep{woosley1993:apj405,macfadyen1999:apj524}, the currently favored progenitor model for long duration gamma-ray bursts. The resemblance between the supernova spectrum distilled from the GRB\,030329 afterglow and that of the energetic type Ic supernova SN1998bw (associated with GRB\,980425, \citealt{galama1998:nature395}) provides strong support for the core collapse of massive stars as the cause for GRBs \citep{hjorth2003:nature423,stanek2003:apj519}.

Several authors have modeled the broad-band afterglow behavior with a standard fireball model for the afterglow. A first approximation shows excess flux (on top of the already bumpy light curve) after the first few days, most noticeable at the lower frequencies. \citet{willingale2004:mnras349} attribute the excess flux to the underlying supernova, but most authors (e.g. \citet{berger2003:nature,sheth2003:apj595,tiengo2004:aap423}) prefer a two-component jet model, where a slower jet is responsible for the extra emission appearing at optical wavelengths around 10 days after the burst. Even later time observations show a likely transition to the non-relativistic regime, estimated around 40 - 50 days after the burst (\citet{tiengo2004:aap423,frail2005:apj619}).

Here we describe our radio monitoring campaign of this extraordinarily bright afterglow with the Westerbork Synthesis Radio Telescope (WSRT) in the centimeter waveband. In Section \ref{SECTION:GRB030329-REDUCTION} we describe the data we obtained. In Section \ref{SECTION:GRB030329-MODELING} we apply an afterglow model to the data, and in Section \ref{SECTION:GRB030329-DISCUSSION} we compare our results with radio data obtained by other groups. Finally, in Section \ref{SECTION:GRB030329-CONCLUSIONS}, we summarize our findings and draw our conclusions.

%
%__________________________________________________________________

\section{Data reduction and analysis \label{SECTION:GRB030329-REDUCTION}}

Data were obtained with the WSRT, at 1.4, 2.3 and 4.8 GHz. We used the Multi Frequency Front Ends \citep{tan1991} in combination with the IVC+DZB backend\footnote{See sect. 5.2 at \url{http://www.astron.nl/wsrt/wsrtGuide/node6.html}} in continuum mode, with a bandwidth of 8x20 MHz. Gain and phase calibrations were performed with the calibrator 3C286, though sometimes 3C147 or 3C48 were used. Table \ref{TABLE:OBSERVATIONS} lists the log of the observations, all done in 2003. VLBI observations prevented us from using the WSRT in the second half of May, and observations were resumed in June, mostly at 4.8 GHz. At 2.3 and 1.4 GHz the observations suffered from confusion from nearby bright sources, causing the noise to be at least a factor of two above the  theoretical limit.

\placetable{TABLE:OBSERVATIONS}
\begin{deluxetable*}{lccccc}
\tablecaption{Log of the WSRT observations of GRB030329 in 2003 \label{TABLE:OBSERVATIONS}}
\tablehead{
\colhead{observing dates} &
\colhead{$\Delta t$\tablenotemark{a}} &
\colhead{integration time} &
\colhead{frequency} &
\colhead{flux} &
\colhead{error flux} \\
\colhead{} &
\colhead{(days)} &
\colhead{(hours)} &
\colhead{(GHz)} &
\colhead{(mJy)} &
\colhead{(mJy)} \\
}
\startdata
Mar 30.658 - 31.138     &  1.414  & 3.3  & 1.4   & 0.21 &  0.07 \\
Mar 30.674 - 31.148     &  1.427  & 3.6  & 2.3   & 0.28 &  0.05 \\
Mar 30.691 - 31.157     &  1.440  & 3.6  & 4.8   & 1.05 &  0.03 \\
Mar 31.655 - Apr 1.110  &  2.399  & 3.6  & 4.8   & 5.98 &  0.03 \\
Mar 31.672 - Apr 1.126  &  2.415  & 3.6  & 2.3   & 2.17 &  0.05 \\
Mar 31.688 - Apr 1.142  &  2.431  & 3.3  & 1.4   & 0.63 &  0.04 \\
Apr 2.650  - 3.088      &  4.385  & 4.0  & 4.8   & 3.64 &  0.04 \\
Apr 2.694  - 3.132      &  4.429  & 4.0  & 2.3   & 0.79 &  0.05 \\
Apr 2.738  - 3.149      &  4.459  & 3.3  & 1.4   & 0.43 &  0.15 \\
Apr 4.644  - 5.083      &  6.380  & 4.0  & 4.8   & 4.89 &  0.04 \\
Apr 4.688  - 5.127      &  6.424  & 4.0  & 2.3   & 1.21 &  0.06 \\
Apr 4.732  - 5.143      &  6.454  & 3.3  & 1.4   & 0.37 &  0.04 \\
Apr 5.641  - 5.818      &  7.246  & 4.2  & 2.3   & 1.00 &  0.07 \\
Apr 7.072  - 7.138      &  8.621  & 1.6  & 4.8   & 4.21 &  0.06 \\
Apr 7.636  - 8.135      &  9.401  & 12.0 & 4.8   & 3.96 &  0.04 \\
Apr 11.625 - 11.693     & 13.175  & 1.6  & 4.8   & 6.42 &  0.05 \\
Apr 12.622 - 12.806     & 14.230  & 2.5  & 4.8   & 5.25 &  0.05 \\
Apr 12.659 - 12.843     & 14.267  & 2.5  & 2.3   & 0.26 &  0.06 \\
Apr 18.606 - 18.908     & 20.273  & 7.0  & 4.8   & 6.96 &  0.05 \\
Apr 20.600 - 21.055     & 22.344  & 3.3  & 4.8   & 6.50 &  0.04 \\
Apr 20.617 - 21.072     & 22.360  & 3.3  & 2.3   & 1.21 &  0.06 \\
Apr 20.633 - 21.087     & 22.376  & 3.3  & 1.4   & 0.64 &  0.08 \\    
Apr 27.583 - 28.062     & 29.338  & 3.9  & 4.8   & 9.17 &  0.04 \\
Apr 27.606 - 28.078     & 29.357  & 5.5  & 2.3   & 2.32 &  0.04 \\
May 2.568  - 2.816      & 34.208  & 6.0  & 2.3   & 3.04 &  0.06 \\
May 3.732  - 3.955      & 37.360  & 5.3  & 4.8   & 8.90 &  0.04 \\
May 9.783  - 9.811      & 41.313  & 0.7  & 4.8   & 8.05 &  0.10 \\
May 9.813  - 9.841      & 41.343  & 0.7  & 2.3   & 3.11 &  0.18 \\
Jun 9.714  - 9.963      & 72.354  & 2.5  & 4.8   & 4.59 &  0.04 \\
Jun 9.737  - 9.944      & 72.356  & 2.5  & 2.3   & 3.84 &  0.07 \\
Jun 16.567 - 16.645     & 79.122  & 1.9  & 4.8   & 4.39\tablenotemark{b} &  0.08 \\
Jun 17.750 - 17.792     & 80.287  & 1.0  & 4.8   & 3.79 &  0.08 \\
Jun 18.777 - 18.819     & 81.314  & 1.0  & 4.8   & 3.74 &  0.10 \\
Jun 30.430 - 30.440     & 92.951  & 0.3  & 4.8   & 2.34 &  0.18 \\
Jul 1.652  - 1.743      & 94.214  & 1.0  & 4.8   & 3.82 &  0.10 \\
Jul 2.488  - 2.524      & 95.022  & 0.9  & 4.8   & 2.67\tablenotemark{b} &  0.08 \\
Jul 3.548  - 3.587      & 96.084  & 1.6  & 4.8   & 3.57 &  0.07 \\
Jul 5.892  - 5.933      & 98.429  & 1.0  & 4.8   & 3.09 &  0.04 \\
Jul 19.580 - 19.854     & 112.233 & 6.6  & 4.8   & 2.27 &  0.05 \\
Jul 23.673 - 23.843     & 116.274 & 4.1  & 4.8   & 2.43 &  0.05 \\ 
Jul 29.519 - 29.644     & 122.097 & 3.0  & 2.3   & 3.00 &  0.06 \\
Jul 29.675 - 29.800     & 122.254 & 3.0  & 1.4   & 1.72 &  0.15 \\
Aug 2.546  - 2.595      & 126.086 & 1.2  & 4.8   & 2.13 &  0.07 \\
Aug 2.619  - 2.668      & 126.159 & 1.2  & 2.3   & 0.80\tablenotemark{b} &  0.27 \\
Aug 2.692  - 2.740      & 126.232 & 1.2  & 1.4   & 1.93 &  0.18 \\
Sep 13.535 - 13.701     & 168.134 & 4.0  & 4.8   & 1.28 &  0.04 \\
Sep 16.196 - 16.250     & 170.739 & 1.3  & 1.4   & 2.29 &  0.19 \\
Sep 28.161 - 28.285     & 182.739 & 3.0  & 2.3   & 1.95 &  0.09 \\
Oct 11.291 - 11.458     & 195.891 & 4.0  & 1.4   & 1.40 &  0.21 \\
Nov 29.248 - 29.491     & 244.885 & 5.8  & 1.4   & 1.15 &  0.09 \\
Dec 1.128  - 1.288      & 246.724 & 3.8  & 4.8   & 0.85 &  0.04 \\
\enddata
%  \begin{list}{}{}
  \tablenotetext{a}{In days after the burst. The indicated time is the logarithmic average of the start and end of the integration.}
  \tablenotetext{b}{The flux of surrounding point sources is consistently lower compared to other observations.}
%  \end{list}
\end{deluxetable*}

We checked our results for consistency by measuring the flux of several nearby point sources, which were assumed to be constant. In a few observations, we found these sources to be systematically dimmer, as indicated in the observation log; we therefore suspect that the flux derived in these observations for the afterglow is also below its real value. Although we could in principle scale these fluxes upward, we decided to ignore these observations in our analysis, as the cause of these low flux levels is not clear.

\begin{figure*}
  \begin{center}
    \includegraphics[height=0.55\textheight]{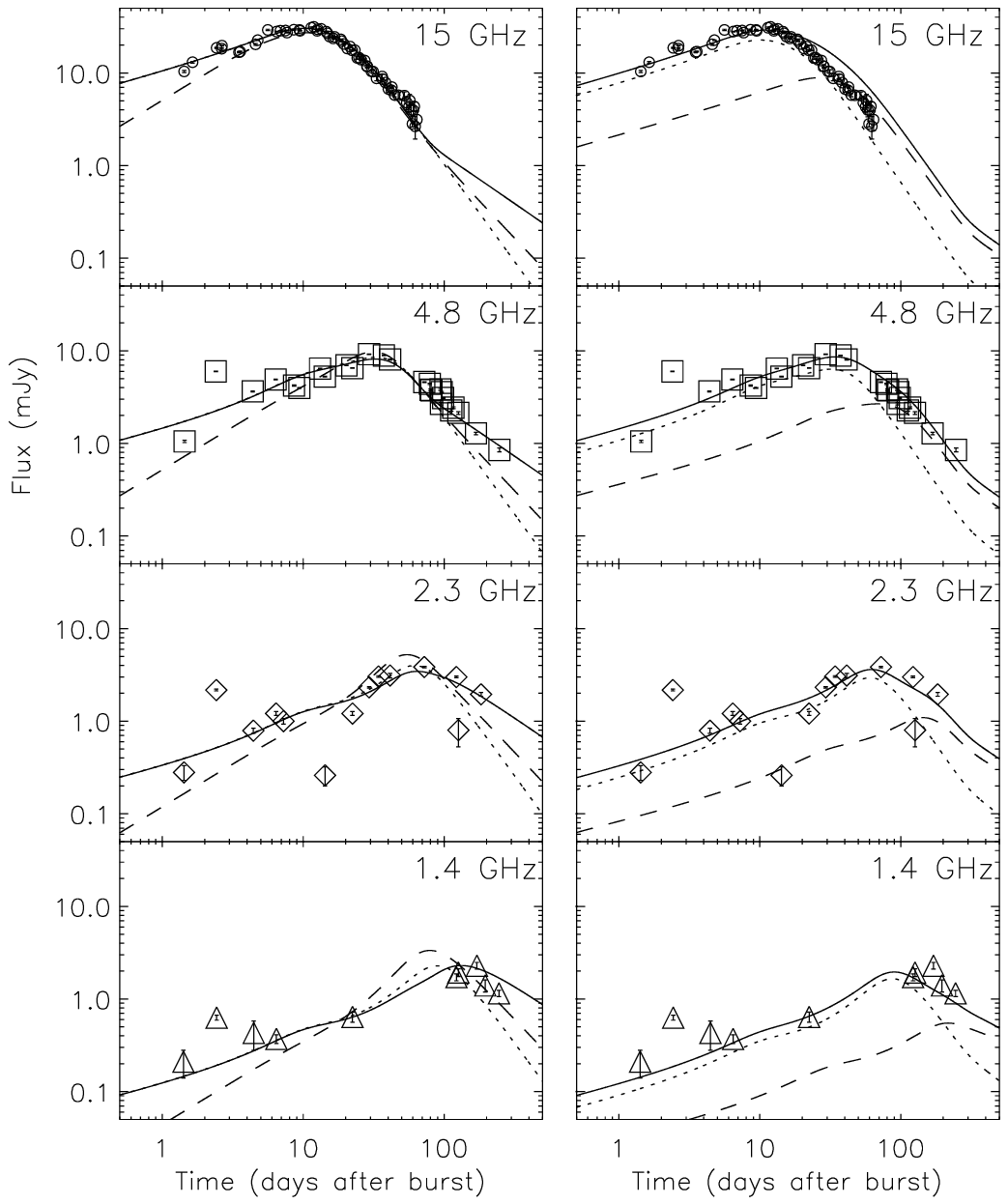}
  \end{center}
  \figcaption{\label{FIGURE:FIT} Eyeball fits to the 4.8, 2.3 and 1.4 GHz WSRT data simultaneously, with 15 GHz VLA/ATCA/RT observations from \citet{berger2003:nature} for comparison. Note the large scatter around the model light curves during the first days. 
{\bf Left:} The lines represent models with $\vm\ \simeq 35$ GHz, $\va\ \simeq 13$ GHz and $\Fm\ \simeq 61$ mJy at $\tj = 10$ days. 
The solid line corresponds to a model in which the fireball expands into a homogeneous medium and the non-relativistic phase of the fireball evolution starts after 80 days; 
the dotted line corresponds to the same model but without a non-relativistic phase, so it deviates from the solid line after 50 days; the dashed line corresponds to a model with a non-relativistic phase after 80 days 
and expansion of the fireball into a massive stellar wind. The peak frequency falls below the self absorption frequency at 17 days. From then on, the maximum of a light curve at a given wavelength marks the passing of the self absorption frequency. 
{\bf Right:} A two-component fit to the data. The first component (dotted line, with a jet break time of 10 days) is responsible for the light curves until 50 days, while the second component (dashed line, with $\tj \simeq 30$ days) accounts for the later peak in the light curves. The combined light curve is shown as the solid line.}
\end{figure*}

\placefigure{FIGURE:FIT}

\begin{figure*}
  \begin{center}
    \includegraphics[height=0.39\textheight]{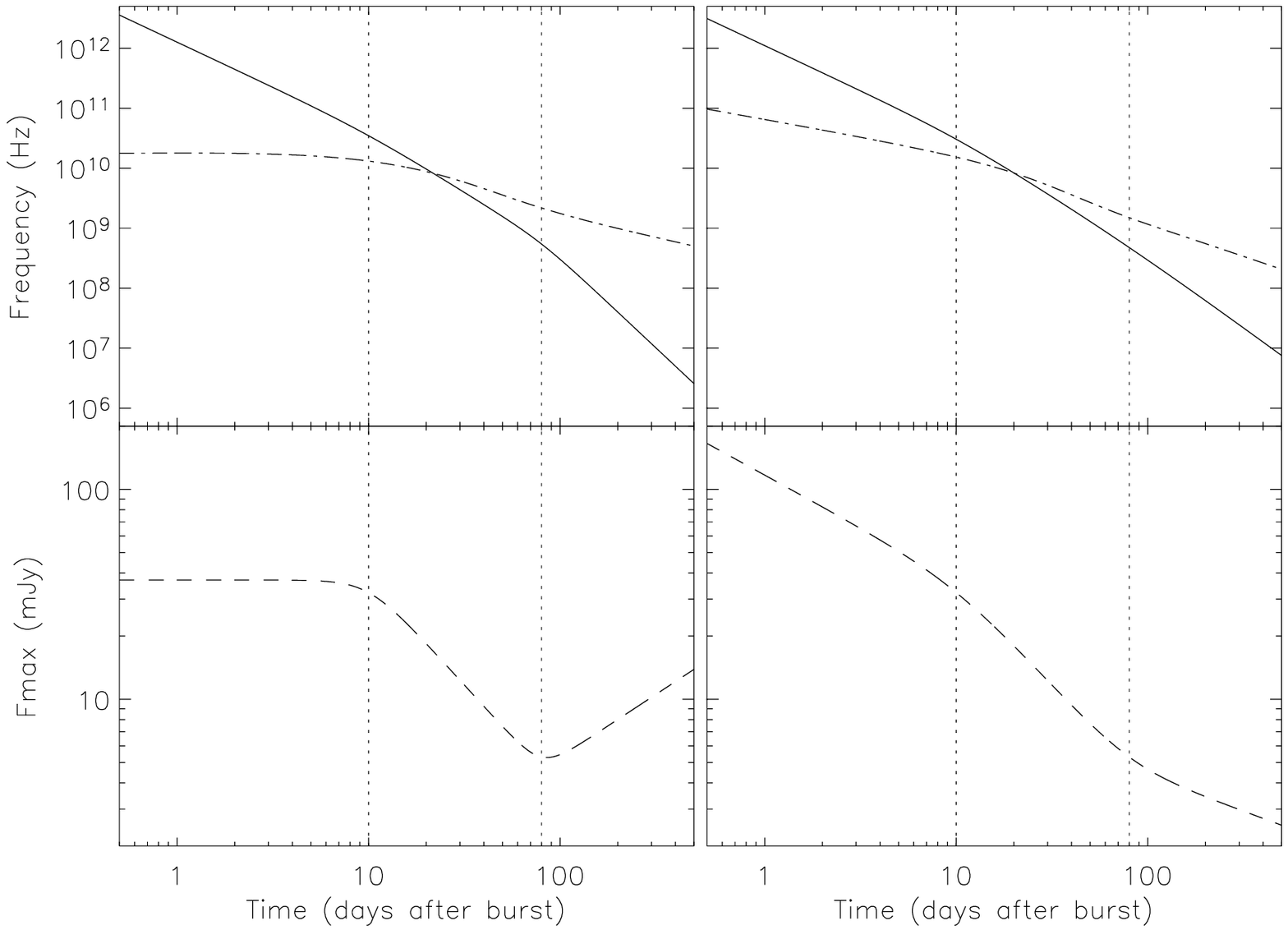}
  \end{center}
  \figcaption{\label{FIGURE:PARAMETERS} The temporal evolution of the electron injection frequency, \vm, the self absorption frequency, \va, and the peak flux, \Fm. 
The upper panels show the evolution of \vm\ (solid line) and \va\ (dash-dotted line), the lower panels the evolution of \Fm. 
The left panels show a model in which the fireball expands into a homogeneous medium, with a jet-break time $\tj=10$ days, 
and the non-relativistic phase of the fireball evolution starts after $\tnr=80$ days. 
The right panels show a model in which the fireball expands into a massive stellar wind, with a jet-break time $\tj=10$ days, 
and the non-relativistic phase of the fireball evolution starts after $\tnr=80$ days. 
The dotted lines show where the breaks in the temporal behavior of the parameters occur at \tj\ and \tnr.}
\end{figure*}

The 1.4, 2.3 and 4.8 GHz light curves are presented in Fig. \ref{FIGURE:FIT}. The general trend of the light curves is that expected for the low-frequency part of a GRB afterglow: as long as the self-absorption frequency, \va, is higher than the observed frequency interval, the light curve rises since the frequency of the minimum electron injection energy \vm\  moves toward lower frequencies. When both \va\ and \vm\ pass the observed frequency interval (not necessarily at the same time), a turn-over in the light curve occurs and the flux falls off steeply. We have listed the precise temporal dependencies of \va, \vm, \vc\ and the peak flux \Fm\ in Table \ref{TABLE:GRB030329-BREAKFREQUENCIES}, for a homogeneous circumburst medium as well as for a massive stellar wind, in which the external density $\rho$ depends on distance $r$ to the center as $\rho \propto r^{-2}$. The evolution in time of \va, \vm\ and \Fm\ is plotted in Fig. \ref{FIGURE:PARAMETERS}. 
Table \ref{TABLE:GRB030329-FLUX-SLOPES} lists the dependencies of the flux on $\nu$ and $t$ (note the change in the spectral index when \vm\ becomes less than \va).

\placetable{TABLE:GRB030329-BREAKFREQUENCIES}

\begin{table*}
\caption{\label{TABLE:GRB030329-BREAKFREQUENCIES} The various temporal dependencies of the break frequencies and peak flux of the afterglow broad band spectrum. 
Before the jet-break time \tj\ and after the non-relativistic timescale \tnr\ different scalings arise from a homogeneous circumburst medium or a stellar wind, 
when the external density $\rho$ depends on distance $r$ to the center as $\rho \propto r^{-2}$. 
Between \tj\ and \tnr\ the external density profile does not influence the scalings.}
\begin{center}
\begin{tabular}{l|lllll}
\tableline
 & \va\ ($\va < \vm\ < \vc$) & \va\ ($\vm < \va\ < \vc$) & \vm & \vc & \Fm \\
\tableline
$t < \tj < \tnr$ (homogeneous)  & $t^{\,0}$     & $t^{\,-(3p+2)/2(p+4)}$ & $t^{\,-3/2}$ & $t^{\,-1/2}$ & $t^{\,0}$ \\
$t < \tj < \tnr$ (stellar wind) & $t^{\,-3/5}$  & $t^{\,-3(p+2)/2(p+4)}$ & $t^{\,-3/2}$ & $t^{\,1/2}$  & $t^{\,-1/2}$ \\
$\tj < t < \tnr$                & $t^{\,-1/5}$  & $t^{\,-2(p+1)/(p+4)}$  & $t^{\,-2}$   & $t^{\,0}$    & $t^{\,-1}$ \\
$\tj < \tnr < t$ (homogeneous)  & $t^{\,6/5}$   & $t^{\,-(3p-2)/(p+4)}$  & $t^{\,-3}$   & $t^{\,-1/5}$ & $t^{\,3/5}$ \\
$\tj < \tnr < t$ (stellar wind) & $t^{\,-2/15}$ & $t^{\,-(7p+6)/3(p+4)}$ & $t^{\,-7/3}$ & $t^{\,1}$    & $t^{\,-1/3}$ \\
\tableline
\end{tabular}
\end{center}
\end{table*}

\placetable{TABLE:GRB030329-FLUX-SLOPES}

\begin{table*}
\caption{\label{TABLE:GRB030329-FLUX-SLOPES} The spectral and temporal flux dependencies in the different regimes of the broad-band afterglow spectrum. }
\begin{center}
\footnotesize
\begin{tabular}{l|llll}
\tableline
$\va < \vm < \vc$ & F ($\nu < \va$) & F ($\va < \nu < \vm$) & F ($\vm < \nu < \vc$) & F ($\vc < \nu$) \\
\tableline
$t < \tj < \tnr$ (homogeneous)  & $\nu^{\,2} \cdot t^{\,1/2}$  & $\nu^{\,1/3} \cdot t^{\,1/2}$  & $\nu^{\,-(p-1)/2} \cdot t^{\,-3(p-1)/4}$   & $\nu^{\,-p/2} \cdot t^{\,-(3p-2)/4}$ \\
$t < \tj < \tnr$ (stellar wind) & $\nu^{\,2} \cdot t^{\,1}$    & $\nu^{\,1/3} \cdot t^{\,0}$    & $\nu^{\,-(p-1)/2} \cdot t^{\,-(3p-1)/4}$   & $\nu^{\,-p/2} \cdot t^{\,-(3p-2)/4}$ \\
$\tj < t < \tnr$                & $\nu^{\,2} \cdot t^{\,0}$    & $\nu^{\,1/3} \cdot t^{\,-1/3}$ & $\nu^{\,-(p-1)/2} \cdot t^{\,-p}$          & $\nu^{\,-p/2} \cdot t^{\,-p}$ \\
$\tj < \tnr < t$ (homogeneous)  & $\nu^{\,2} \cdot t^{\,-2/5}$ & $\nu^{\,1/3} \cdot t^{\,8/5}$  & $\nu^{\,-(p-1)/2} \cdot t^{\,-3(5p-7)/10}$ & $\nu^{\,-p/2} \cdot t^{\,-(3p-4)/2}$ \\
$\tj < \tnr < t$ (stellar wind) & $\nu^{\,2} \cdot t^{\,2/3}$  & $\nu^{\,1/3} \cdot t^{\,4/9}$  & $\nu^{\,-(p-1)/2} \cdot t^{\,-(7p-5)/6}$   & $\nu^{\,-p/2} \cdot t^{\,-(7p-8)/6}$ \\[2mm]
\tableline
$\vm < \va\ < \vc$ & F ($\nu < \vm$) & F ($\vm < \nu < \va$) & F ($\vm < \nu < \vc$) & F ($\vc < \nu$) \\
\tableline
$t < \tj < \tnr$ (homogeneous)  & $\nu^{\,2} \cdot t^{\,1/2}$  & $\nu^{\,5/2} \cdot t^{\,5/4}$   & $\nu^{\,-(p-1)/2} \cdot t^{\,-3(p-1)/4}$   & $\nu^{\,-p/2} \cdot t^{\,-(3p-2)/4}$ \\
$t < \tj < \tnr$ (stellar wind) & $\nu^{\,2} \cdot t^{\,1}$    & $\nu^{\,5/2} \cdot t^{\,7/4}$   & $\nu^{\,-(p-1)/2} \cdot t^{\,-(3p-1)/4}$   & $\nu^{\,-p/2} \cdot t^{\,-(3p-2)/4}$ \\
$\tj < t < \tnr$                & $\nu^{\,2} \cdot t^{\,0}$    & $\nu^{\,5/2} \cdot t^{\,1}$     & $\nu^{\,-(p-1)/2} \cdot t^{\,-p}$          & $\nu^{\,-p/2} \cdot t^{\,-p}$ \\
$\tj < \tnr < t$ (homogeneous)  & $\nu^{\,2} \cdot t^{\,-2/5}$ & $\nu^{\,5/2} \cdot t^{\,11/10}$ & $\nu^{\,-(p-1)/2} \cdot t^{\,-3(5p-7)/10}$ & $\nu^{\,-p/2} \cdot t^{\,-(3p-4)/2}$ \\
$\tj < \tnr < t$ (stellar wind) & $\nu^{\,2} \cdot t^{\,2/3}$  & $\nu^{\,5/2} \cdot t^{\,11/6}$  & $\nu^{\,-(p-1)/2} \cdot t^{\,-(7p-5)/6}$   & $\nu^{\,-p/2} \cdot t^{\,-(7p-8)/6}$ \\
\tableline
\end{tabular}
\end{center}
\end{table*}

\placefigure{FIGURE:PARAMETERS}

\section{Applying the fireball model to the data\label{SECTION:GRB030329-MODELING}}

We have modeled the data simultaneously in time and frequency, using a general broad-band afterglow model which includes a jet break and a transition to non-relativistic expansion of the fireball. The several power-law segments of the broad-band spectrum were connected smoothly in a way described in Appendix \ref{SECTION:APPENDIX}. Because \Fm, \va, \vm\ and \vc\ are functions of time, we need to quote them at some fixed moment, for which we choose the jet-break time. At earlier and later times the characteristic frequencies and the peak flux evolve in time according to Table \ref{TABLE:GRB030329-BREAKFREQUENCIES}. The resulting light curves are given in Table \ref{TABLE:GRB030329-FLUX-SLOPES}. The transitions between the different regimes, marked by the jet-break time $\tj$ and the time $\tnr$ at which the fireball becomes non-relativistic, are treated as smoothly broken power laws as described in Appendix \ref{SECTION:APPENDIX}. 

Since our data show a large scatter, especially at early times due to scintillation, where they do not follow a smooth curve, we did not apply a $\chi^2$ fit, but merely tried to obtain a best fit by eye. This ignores the scintillation, and it will also put some more emphasis on the 1.4 GHz light curve: a $\chi^2$ fit tends to follow the 4.8 GHz data points since they are more numerous, and will hence ignore the global trend seen in the 1.4 GHz light curve. 
The result of such an eye-ball broad-band fit is shown in Fig. \ref{FIGURE:FIT}. Note that the scintillation amplitude is largest in the first few days, after which it quickly declines (the large decrease in the 2.3 GHz light curve around 20 days could be an artifact in the data, since it does not show up in the other light curves). 

In this way, we obtained a value of 35 GHz for the electron injection frequency, \vm, and a value of 13 GHz for the self absorption frequency, \va, at 10 days after the burst, which is the jet-break time. The flux at \vm\ is about 61 mJy at that time. We find the electron index to be $p = 2.1$. The value of \vc\ can only be determined with observations at higher frequencies; our dataset indicates that $\vc\ \ga 10^{12}$ Hz, which is in agreement with the findings by \citet{smith2005:aasubm}, who find that the rapid fall in their $350\,\mbox{GHz}$ light curve can be attributed to the cooling frequency passing through their observing band.
These results are somewhat at odds with the findings by \citet{berger2003:nature}. They obtain higher values for the characteristic frequencies \va\ ($19$ GHz) and \vm\ ($43$ GHz), and the peak flux \Fm\ ($96$ mJy) at the jet-break time $t_{\mathrm{j}} \simeq 9.8$ days. To investigate this further, we performed a fit which includes their data, as well as radio data from \citet{sheth2003:apj595}. This fit gives similar results to those obtained from our previous fit. 

From the obtained characteristic frequencies and the peak flux we find an isotropic energy $\Eiso\ \simeq 4.0\cdot 10^{51}\vcg^{1/4} \,\mathrm{erg}$, a circumburst density $n \simeq 0.56\vcg^{3/4}\,\mathrm{cm}^{-3}$, and the fractions of energy in the electrons $\ee$ and magnetic field $\eb$ of $0.25\vcg^{1/4}$ and $0.49\vcg^{-5/4}$ respectively. The cooling frequency can not be determined from the radio observations at centimeter wavelengths, but we take $\vc\ \equiv 10^{13} \vcg $ to compare our results with \citet{berger2003:nature}. The opening angle of the jet can be found to be $\theta_{\mathrm{j}} \simeq 0.38\vcg^{-1/2}\,\mathrm{rad}$ ($22^{\circ}$) from the jet-break time of 10 days, which gives a beaming-corrected energy of $\Ecor \simeq 2.9\cdot 10^{50}\vcg^{-3/4}\,\mathrm{erg}$. This energy is comparable to \citet{berger2003:nature}, the circumburst medium density we find is smaller, but $\ee$ and $\
eb$ are larger.

At the turnover in the late-time light curves at 1.4, 2.3 and 4.8 GHz we find an excess in flux compared to the model. We present two possible explanations for this behavior. The first one is a non-relativistic phase after $\tnr \simeq 80$ days. Table \ref{TABLE:GRB030329-FLUX-SLOPES} shows that the light curves flatten in the transition to a non-relativistic phase. In Fig. \ref{FIGURE:FIT} one can see that this extension of the model fits the data at 2.3 and 4.8 GHz well when one assumes a homogeneous external medium. However, the flux at 1.4 GHz is overestimated at the latest times in this case (note that the symbols are larger than the error bars). The model with an external density gradient and a non-relativistic phase gives a better result at 1.4 GHz, but underestimates the flux at higher frequencies. 

The second explanation of the late-time flux excess is an extra component, which consists of an afterglow with a jet-break time later than 10 days. The resultant fit is also shown in Fig. \ref{FIGURE:FIT}. For the first component, which produces the main flux at higher frequencies, the parameters are set as before except for $\Fm\ \simeq 48$ mJy. The second component has $\tj \simeq 30$ days, $\vm\ \simeq 20$ GHz, $\va\ \simeq 10$ GHz and $\Fm\ \simeq 16$ mJy at the jet-break time; so at $t=10$ days the second component has $\vm\ \simeq 35$ GHz, $\va\ \simeq 13$ GHz and $\Fm\ \simeq 48$ mJy. The electron index $p \simeq 2.2$ for both jets. The physical parameters we derive from these characteristic frequencies and peak flux are $n \simeq 0.82\vcg^{3/4}\,\mathrm{cm}^{-3}$, $\ee\ \simeq 0.28\vcg^{1/4}$ and $\eb\ \simeq 0.43\vcg^{-5/4}$ for both components; for the first component we find $\Eiso\ \simeq 4.0\cdot 10^{51}\vcg^{1/4} \,\mathrm{erg}$, $\theta_{\mathrm{j}} \simeq 0.42\vcg^{-1/2}\,\mathrm{rad}$ ($24^{\circ}$) and $\Ecor \simeq 2.4\cdot 10^{50}\vcg^{-3/4}\,\mathrm{erg}$, while for the second component $\Eiso\ \simeq 9.1\cdot 10^{50}\vcg^{1/4} \,\mathrm{erg}$, $\theta_{\mathrm{j}} \simeq 0.73\vcg^{-1/2}\,\mathrm{rad}$ ($42^{\circ}$) and $\Ecor \simeq 2.4\cdot 10^{50}\vcg^{-3/4}\,\mathrm{erg}$. The data are well fitted by this two-component jet model. However, data at higher radio frequencies from \citet{berger2003:nature} can not be fitted well in this model.

\section{Discussion\label{SECTION:GRB030329-DISCUSSION}}

A similar procedure of fitting two components with different jet breaks was applied by \citet{berger2003:nature} to explain the break in the early-time optical (and X-ray) light curve. The underlying mechanism involves two jetted outflows, one with a small opening angle and a high Lorentz factor that produces the early-time light curve (with $\tj\simeq 0.5$ days), and a jet with a larger opening angle and lower Lorentz factor that carries the bulk of the energy and produces the later-time light curve (with $\tj\simeq 10$ days). 
The WSRT observations at $2.4$ days after the burst have values for the flux that are well above the theoretical curves (see Fig. \ref{FIGURE:FIT}). We investigated the possibility that these are signatures of the jet that produces the early-time optical light curve. However, with the constraints on the parameters from the optical and X-ray observations, it is not possible to fit the early radio observations with this jet with a jet-break time of $0.5$ days. A better explanation for these observations is scintillation.

From the result of our two-component model fit, we can conclude that, besides the jets with $\tj\simeq 0.5$ and $\tj\simeq 10$, another jet is present with even larger opening angle, that powers the late-time ($t > 50$ days) light curve and is therefore best visible at the very low frequencies observed here. However, it may be that the total jet (which possibly includes the first narrow jet as well) is structured \citep[e.g.][]{meszaros1998:apj499,rossi2002:mnras332} and that the Lorentz factor $\Gamma$ decreases toward the edge of the jet-cone. Alternatively, the outflow consists of a layered jet, where shocks with lower $\Gamma$ follow the faster ones as they run into the surrounding medium. In both cases, one expects that the low Lorentz factors dominate at low frequencies and late times, and that the jet break occurs later at progressively lower frequencies. 

The multiple component model is certainly not satisfactory: it does not give a good fit to the data at radio frequencies above 4.8 GHz at late times. 
Our model in which a transition to a non-relativistic phase of the fireball occurs after 80 days gives a better broadband radio fit except for the data at 1.4 GHz. 
This transition to a non-relativistic phase is also seen in VLA radio observations by \citet{frail2005:apj619} and at X-ray frequencies by \citet{tiengo2004:aap423}, 
although their estimate for the time at which this transition occurs is lower than ours, i.e. $\sim 50$ and $\sim 44$ days respectively. 
Our low frequency radio data can not be fitted well by applying this low value for the non-relativistic transition. 

Although the model in which a transition to a non-relativistic phase of the fireball occurs, gives the best broadband radio fit, the value of $\tnr$ we find is much lower than theoretical estimates done by \citet{granot2005:apj618} and \citet{oren2004:mnras353}, based on determinations of the evolution of the image size of GRB\,030329 (\citet{taylor2005:apj622}). We estimate $\tnr\ \simeq 209 (\frac{\Ecor}{10^{51}n})^{1/3}\simeq 168\vcg^{-1/6}\,\mathrm{days}$, which is a factor of $2$ higher than the $80$ days we get from modeling the centimeter light curves. This discrepancy could be solved by fitting the broadband afterglow light curves of GRB\,030329 simultaneously with the evolution of its image size.

\section{Summary and conclusions\label{SECTION:GRB030329-CONCLUSIONS}}

Our data confirm the picture of a second jet in the afterglow of GRB\,030329, that manifests itself around $\tj = 10$ days. However, the flux level around the time when the low frequency light curves peak is higher than that predicted by the two-component afterglow model \citep[cf.][]{berger2003:nature}. Adding a third component with a later \tj, we can account for this excess flux. Taking into account the early jet break, seen most clearly at optical wavelengths, we suggest that one is actually seeing the result of several blast waves with a range in Lorentz factors \citep{rees1998:apj496,granot2003:nature426}, something which comes quite naturally in the collapsar model for GRBs \citep{macfadyen2001:apj550,ramirezruiz2002:mnras337,zhang2003:apj586}, and was already suggested by \citet{sheth2003:apj595}. However, their high frequency data was unable to distinguish jet breaks at later times. Our later time low frequency radio data show such a late-time jet break, corresponding to a lower Lorentz factor, and therefore support a layered or structured jet for the afterglow of GRB\,030329. 

An alternative explanation is the transition to a non-relativistic phase of the fireball. 
This model gives a good fit to the data at 2.3 and 4.8 GHz, but overestimates the flux at 1.4 GHz. 
This overestimation can be caused by the method of smoothly broken power laws as described in Appendix \ref{SECTION:APPENDIX}. 
We did not take into account the jet that is pointing away from us, and this can possibly give extra flux at late times when the jet becomes non-relativistic and spherical. 
Continuation of observations at late times at low radio frequencies and more detailed physical models can diagnose the cause of this discrepancy more closely.

\acknowledgements
We greatly appreciate the support from the WSRT staff in their help with scheduling these observations as efficiently as possible. We thank Ed van den Heuvel for useful discussions. ER acknowledges support from NWO grant nr. 614-51-003. ER would like to thank the hospitality of the observatory of Padua. The authors acknowledge benefits from collaboration within the Research Training Network "Gamma-Ray Bursts: An Enigma and a Tool", funded by the EU under contract HPRN-CT-2002-00294.

\appendix
\section{Broadband spectrum and light curve modeling\label{SECTION:APPENDIX}}
The dominating radiation mechanism for GRB afterglows is synchrotron emission. The broadband synchrotron spectrum is determined by the peak flux and three break frequencies, namely the synchrotron self-absorption frequency, \va, the frequency that corresponds to the minimal energy in the electron energy distribution, \vm, and the cooling frequency, \vc, that corresponds to electrons that lose their energy quickly by radiation. The time evolution of these four parameters gives the evolution of the spectrum and thus light curves at all observing frequencies. 

The relativistic electrons emitting the synchrotron radiation are accelerated at the shock front to a power-law distribution of energies, 
$N(\gamma_{e})d\gamma_{e}\propto\gamma_{e}^{-p}d\gamma_{e}$ for $\gamma_{e}\geq\gamma_{m}$. 
For electrons that cool on a timescale smaller than the dynamical timescale, the energy distribution is steeper, $N(\gamma_{e})d\gamma_{e}\propto\gamma_{e}^{-p-1}d\gamma_{e}$. The broadband energy spectrum is found by connecting these two energy distributions at the cooling Lorentz factor $\gamma_c$, and then integrating the single-electron synchrotron spectrum over the distribution function. Synchrotron self-absorption is taken into account by calculating the flux $F_{\nu}$ as follows:
\begin{equation}F_{\nu}=\frac{j_{\nu}}{D^2}(\frac{\alpha_{\nu}}{\alpha_{\nu_a}})^{-1}[1-\exp{(-\frac{\alpha_{\nu}}{\alpha_{\nu_a}})}]\,,\end{equation}
with $j_{\nu}$ the emission coefficient, $\alpha_{\nu}$ the absorption coefficient, and $D$ the distance. 
A detailed description of our modeling is presented in \citet{vanderhorst2005:inprep}.

The evolution of the characteristic frequencies and the peak flux is given in Table \ref{TABLE:GRB030329-BREAKFREQUENCIES}. We assume that after the jet-break time the jet spreads sideways (\citet{rhoads1999:apj525}), until it becomes spherical approximately at the same time the fireball becomes non-relativistic and enters the Sedov-Von Neumann-Taylor phase of the evolution. The transitions between the different regimes, marked by the jet-break time $\tj$ and the time $\tnr$ at which the fireball becomes non-relativistic, are treated as smoothly broken power laws. We introduce a smoothening parameter $s$ and take \vm\ and \vc\ as examples:
\begin{equation}\vm(t)=\vm(\tj)\cdot \left[\left(\frac{t}{\tj}\right)^{3/2\cdot s}+\left(\frac{t}{\tj}\right)^{2\cdot s}
+\left(\frac{\tnr}{\tj}\right)^{2\cdot s}\cdot \left(\frac{t}{\tnr}\right)^{3\cdot s}\right]^{\,-1/s}\,,\end{equation}
\begin{equation}\vc(t)=\vc(\tj)\cdot \left[\left(\frac{t}{\tj}\right)^{-1/2\cdot s}+\left(1+\left(\frac{t}{\tnr}\right)^{1/5\cdot s}\right)^{-1}\right]^{\,1/s}\,.\end{equation}
Expressions for \va\ and \Fm\ are similar. 

We choose to have a smoothening parameter of $s=5$ for every transition. However, these transitions will probably be different and this can only be accounted for in detailed hydrodynamical modeling of the fireball. So this smoothening parameter gives a an uncertainty which may account for the discrepancy seen at 1.4 GHz in the case of one jet with a transition to the non-relativistic phase at $\tnr\simeq 80$ days.

\bibliographystyle{apj}
%\bibliography{references}

\end{document}